\documentclass{jpsj2}
\title{High Field ESR and Magnetization of the Triangular Lattice Antiferromagnet NiGa$_2$S$_4$}

\author{Hironori \textsc{Yamaguchi}\thanks{E-mail: yamaguchi@mag.cqst.osaka-u.ac.jp}, Shojiro \textsc{Kimura}, Masayuki \textsc{Hagiwara}, Yusuke \textsc{Nambu}$^{1, 2}$,\\
Satoru \textsc{Nakatsuji}$^{3}$, Yoshiteru \textsc{Maeno}$^{4}$, Akira \textsc{Matsuo}$^{3}$, and Koichi \textsc{Kindo}$^{3}$}
\inst{KYOKUGEN, Osaka University, 1-3 Machikaneyama, Toyonaka, Osaka 560-8531 \\
$^{1}$Department of Physics and Astronomy, The Johns Hopkins University, Baltimore, Maryland 21218, USA\\
$^{2}$NIST Center for Neutron Research, Gaithersburg, Maryland 20899, USA\\
$^{3}$Institute for Solid State Physics, University of Tokyo, 5-1-5 Kashiwanoha, Kashiwa, Chiba 277-8581 \\
$^{4}$Departmen of Physics, Kyoto University, Kyoto, 606-8502}

\abst{We report the experimental and the analytical results of electron spin resonance (ESR) and magnetization in high magnetic fields up to about 68 T of the quasi two-dimensional triangular lattice antiferromagnet NiGa$_2$S$_4$. From the temperature evolution of the ESR absorption linewidth, we find a distinct disturbing of the development of the spin correlation by $Z_2$-vortices between 23 K and 8.5 K. Below $T_{\rm{v}}=8.5$ K, spin-wave calculations based on a 57$^{\circ}$ spiral spin order well explains the frequency dependence of the ESR resonance fields and high field magnetization processes for $H$$\parallel$$c$ and $H$$\perp$$c$, although the magnetization for $H$$\perp$$c$ at high fields is different from the calculated one. Furthermore, we explain the field independent specific heat with $T^2$-dependence by the same spin-wave calculation, but the magnitude of the specific heat is much less than the observed one. Accordingly, these results suggest the occurrence of a $Z_2$ vortex-induced topological transition at $T_{\rm{v}}$ and may indicate quantum effects beyond the descriptions based on the above classical spin models.}   

\kword{high field ESR, high field magnetization, triangular lattice, frustrate magnetism, NiGa$_2$S$_4$}

\begin{document}
\maketitle

\section{Introduction} 
Two-dimensional (2D) triangular lattice antiferromagnets (TLAFM) have attracted considerable attention over the three decades, because some of them are believed to have a novel ground state such as a spin liquid caused by strong frustration~\cite{anderson}. Recently, a new quasi-2D $S$=1 TLAFM NiGa$_2$S$_4$ has attracted a substantial interest as a candidate of such a spin liquid state~\cite{science}. It has slabs consisting of two GaS layers and a NiS$_2$ one separated by a van der Waals gap as shown in Fig. 1(a). Despite strong antiferromagnetic (AF) interactions with Weiss temperature ${\theta}\rm{_w}\simeq$$-80$ K, no long-range order (LRO) down to 0.08 K was observed~\cite{NMR}. Some noteworthy characteristic features were observed in neutron diffraction, specific heat, and magnetic susceptibility measurements~\cite{science}. The neutron diffraction measurements unveil the development of a short-range correlations with the wavevector $\mib{q}$ = (0.158, 0.158, 0), which correspond to a short-range order (SRO) with about 57$^{\circ}$ spiral spin structure as shown in Fig. 1(b). The correlation length of the SRO is only about 2.5 nm (about seven lattice spacings) even at 1.5 K, and the origin of the absence of LRO has not been clarified yet. The magnetic susceptibility ($M/H$) measured at 0.01 T exhibits a weak peak at about 8.5 K and a finite value at $T\rightarrow0$ K, indicating the existence of gapless excitations. The magnetic specific heat shows broad peaks at about 10 K and 80 K, and $T$$^2$-dependence with no field dependence up to 7 T below 10 K. From the $T$$^2$-dependent specific heat, one may associate the existence of a $k$-linear excitation in a 2D system attributed to a Nambu-Goldstone mode resulting from a breaking of a continuous symmetry. These curious experimental results have stimulated further theoretical and experimental studies on this compound.~\cite{NMR,AFQ,newquadra,kawamuraNiGaS,j3riron,takubo,tamurakun,fePRL,single,nambusan,uSR,ESR} A theoretical analysis by taking bilinear-biquadratic exchange interactions into consideration seemed to account for some of such peculiar experimental results~\cite{AFQ,newquadra,kawamuraNiGaS}. Strong third-neighbor and negligible second-neighbor exchange interactions are suggested by analyses of a photoemission spectroscopy~\cite{takubo} and $ab$ $initio$ calculations by a density functional theory (DFT)~\cite{j3riron}. Taking the strong third-neighbor exchange interactions into account, a finite-temperature phase transition is discussed using a Monte Carlo method~\cite{tamurakun}. Recent NMR and $\mu$SR measurements evidenced the occurrence of a novel state where the spins have a MHz-scale dynamics through a weak freezing below about 10 K~\cite{NMR,uSR}, and impurity effect studies of Ni$_{1-x}M_x$Ga$_2$S$_4$ ($M$ = Mn, Fe, Co and Zn) suggested that the integer size of the Heisenberg spins is important to stabilize the 2D coherent behavior in the frozen spin-disordered state~\cite{nambusan}.

In this paper, we report the experimental and analytical results of electron spin resonance (ESR) and magnetization measurements on single crystals of NiGa$_2$S$_4$, which were performed to clarify the spin dynamics in more details than those in our previous short report~\cite{ESR}. We found peculiar dynamics of $Z_2$ vortex-induced effect on the temperature dependence of ESR absorption linewidth. The frequency dependence of the ESR resonance fields, high field magnetizations and low-temperature specific heat below 8.5 K are explained to a some extent by a conventional spin-wave theory, which is consistent with spin-wave excitations expected in the $Z_2$ vortex-induced phase. These results suggest an occurrence of a $Z_2$ vortex-induced topological transition at 8.5 K. However, some results are quantitatively different from the calculated ones based on a classical spin model. In particular, the calculated specific
heat value below about 10 K is largely different from the observed one, suggesting an additional quantum effects such as quadrupolar correllations.

\section{Materials and Methods}

\subsection{Crystal structure}
First we summarize the structural properties of NiGa$_2$S$_4$. Single crystals were grown by chemical-vapour transport using iodine. The details of crystal growing was reported in Ref. \ref{single}. NiGa$_2$S$_4$ belongs to a trigonal system with $P\bar{3}m1$ symmetry and the lattice constants $a$ = $b$ = 0.3624 nm and $c$ = 1.1999 nm, and has a layered unit consisting of central magnetic layer of edge-sharing NiS$_6$ octahedra and nonmagnetic two layers of GaS$_4$ tetrahedra, in which the magnetic layer is sandwiched, as shown in Fig. 1(a). These layered units are separated from each other along the $c$-axis by a van der Waals gap, resulting in highly 2D system with a weak ferromagnetic interlayer interaction~\cite{science}. In the NiS$_2$ layers, Ni$^{2+}$ (3$d$$^8$) ions with an electronic configuration $t$$^6_{2g}$$e$$^2_g$ ($S$ = 1) form the exact TLAFM as shown in Fig. 1(b).

\subsection{Experimental methods}
ESR measurements in pulsed magnetic fields up to about 53 T were carried out at temperatures between 1.3 K and about 100 K using our pulsed field ESR apparatus equipped with a non-destructive pulse magnet, a far-infrared laser (Edinburgh Instruments, U.K.) which cover the frequencies up to about 1.4 THz. We also performed ESR measurements at 1.4 K in static magnetic fields up to 14 T utilizing a superconducting magnet (Oxford Instruments, U.K.) and a vector network analyzer with some extension (ABmm, France) for the frequencies up to about 500 GHz. 
High-field magnetization measurements in pulsed magnetic fields up to about 68 T were carried out with a non-destructive pulse magnet at 1.3 K at ISSP in The University of Tokyo. The magnetization was measured by an induction method using a pick-up coil. In the experiments, external fields were applied parallel ($H$ $\perp$ $c$) and perpendicular ($H$ $\parallel$ $c$) to the 2D planes.

\section{Experimental Results}

\subsection{ High field ESR}
Observed ESR absorption spectra are displayed in figures 2(a)-(d). Figures 2(a) and 2(b) show the temperature dependence of the ESR absorption spectra at 584.8 GHz for $H$ $\parallel$ $c$ and $H$ $\perp$ $c$, respectively. We observed a strong resonance signal, whose linewidth is extremely broad, and a weak resonance signal for both directions. The resonance fields are indicated by the arrows. Since the weak resonance signals are almost temperature independent, it is considered to arise from paramagnetic resonances caused probably by imperfections in the lattices. The temperature dependences of the resonance field and the absorption linewidth of the strong signal are shown in Figs. 3 and 4, respectively. The resonance field shifts to a lower field with decreasing temperature for both directions below about 30 K. The inset of Fig. 3 shows the $g$-values converted from the resonance fields. The absorption linewidth increases as the temperature is lowered from about 80 K, corresponding to the absolute value of the Weiss temperature $|{\theta}\rm{_w}|$. The temperature dependence of the linewidth shows a bending at about 23 K, below which the increasing tendency with decreasing temperature becomes gradual. Below about 8.5 K, the linewidth shows a gradual decrease as the temperature is lowered.  

The frequency dependences of the ESR absorption spectra at 1.3 K for $H$ $\parallel$ $c$ and $H$ $\perp$ $c$ are presented in Figs. 2(c) and 2(d), respectively. The resonance fields are indicated by the arrows. All the resonance fields are plotted in the frequency-field plane as seen in Fig. 5. The resonance modes from the strong signals have a zero-field gap and come close to the paramagnetic resonance line in the higher field for both $H$ $\parallel$ $c$ and $H$ $\perp$ $c$, and such behavior of the ESR modes is similar to that observed in some spiral antiferromagnets with easy-plane anisotropy~\cite{spiralmode}. Since the resonance frequencies of the weak resonance signals are almost proportional to the external fields, they can be considered to arise from the paramagnetic resonance as mentioned in the previous paragraph.

\subsection{High field magnetization}
Figures 6(a) and 6(b) indicate magnetization curves at 1.3 K in pulsed magnetic fields for $H$ $\parallel$ $c$ and $H$ $\perp$ $c$, respectively. For the former, the magnetization increases linearly with increasing field for the whole range of applied magnetic field, while, for the latter, it deviates from a linear increase shown by the broken line above about 30 T. In addition, we observed a small anomaly at about 44 T as shown in the inset of Fig 6(b).

\section{Analyses and Discussion} 

\subsection{Temperature dependence of the ESR linewidth}
From the temperature dependence of the ESR linewidth as shown in Fig. 4, we divide the temperature region into three areas, (i) above 23 K, (ii) between 23 K and 8.5 K, and (iii) below 8.5 K. In the high-temperature region (i), we analyze the temperature dependence of the ESR linewidth in terms of (a) a common expression of the 2D critical behavior and (b) a non-linear sigma (NL${\sigma}$) model~\cite{NLsigma} associated with the AF LRO in a 2D Heisenberg AFM.
In the case (a), the temperature dependence of the linewidth is expressed by the following equation, 
\begin{align}
\Delta H_{1/2} \propto   (T-T_{\rm{N}})^{-p},\label{p}
\end{align} 
where $T_{\rm{N}}$ is the AF long-range ordering temperature, $p$ is a critical exponent, which reflects the anisotropy and the dimensionality of the system. Based on Kawasaki's dynamic equations, the critical exponent of the 2D AFM is given as $p=(3-2\eta )\nu,\label{pnosiki}$ 
where $\nu$ is the critical exponent for the inverse of the spin correlation length, $\eta $ the critical exponent related to the static spin correlation which is assumed to obey the form, ${\textless}S_q^zS_q^z{\textgreater}=({\kappa}^2+q^2)^{-1+{\eta}/2}$~\cite{p-riron}. In the case of the 2D Heisenberg AFM, $p$ is evaluated to be about 2.5 from the experimentally obtained $\eta =0.2$ and $\nu=0.95$, and the value $p$ is known to be close to the experimental values in several Heisenberg 2D compounds~\cite{p-bunken}. In NiGa$_2$S$_4$, we expect that the development of the SRO gives the same critical behavior as in such 2D Heisenberg AFMs. Since there is no LRO down to 0.08 K~\cite{NMR}, we assume that $T_{\rm{N}}$ is approximate to zero. Then, we fit Eq. (1) with $p=2.5$ and $T_{\rm{N}}\simeq 0$ to the experimental linewidth and obtain good agreement between them as shown by the dotted line in Fig. 4. 

In the case (b), we use the following relation of the temperature dependence of the ESR linewidth calculated by Chakravarty and Orbach for an $S=1/2$ nearest-neighbor quantum Heisenberg AFM on a square lattice with a small interplanar coupling, 
\begin{align}
\Delta H_{1/2} \propto   {\xi}^{3}\frac{(k_BT/2{\pi}{\rho}\rm{_s})^{5/2}}{(1+k_BT/2{\pi}{\rho}\rm{_s})^{4}},\label{dH_NLS}
\end{align}  
where ${\xi}$ is a spin-correlation length, $k_{\rm{B}}$ is the Boltzmann constant, and  ${\rho}\rm{_s}$ is a spin-stiffness constant~\cite{NLsigma}. The temperature dependence of the spin-correlation length for 2D frustrated quantum antiferromagnets predicted by a NL${\sigma}$ model is given by~\cite{NLsigma_xi}
\begin{align}    {\xi}\propto\frac{1}{\sqrt{T}}\exp(\frac{4{\pi}{\rho}\rm{_s}}{{k\rm{_B}T}}).
\end{align}
The ${\rho}{\rm{_s}}/k_{\rm{B}}$ is evaluated to be about 18 K from a NL${\sigma}$ model analysis of dynamic relaxation rates of $\mu$SR measurements~\cite{uSR}, which is the same value as that evaluated from the specific heat~\cite{science}. We fit Eq. (\ref{dH_NLS}) to the experimental results with this value. The theoretical result in Fig.4 (gray line) indicates a more rapid increase of the linewidth than the experimental one. This disagreement may originate from strong spin frustration as discussed in the study of the nuclear spin-lattice relaxation rate in Ref. \ref{ito}. 
  
Below 23 K, the increase of the linewidth with decreasing temperature becomes gradual, indicating that some effect disturbs the development of the SRO below 23 K. In the temperature region (ii), we examine two types of stable point defect in 2D systems, (c) a $Z_2$ vortex for the Heisenberg TLAFM and (d) a conventional Kosteritz-Thouless (KT) vortex for $XY$ TLAFM.
The $Z_2$ vortex is a topologically stable point defect characterized by a two-valued topological quantum number~\cite{Z2kawamura}. Unlike the KT vortex of spins, it is a vortex of spin chiralities. The vortices form bound pairs at low temperatures and begin to dissociate at a certain critical temperature $T\rm{_V}$ like a KT-type phase transition. Above $T\rm{_V}$, the unbound free vortices are thermally excited, and the spin fluctuates through the spacial passage of the vortices. The effect of the $Z_2$ vortices on the ESR linewidth was discussed on Heisenberg TLAFMs, HCrO$_2$ and LiCrO$_2$~\cite{Z2linewidth}. According to the analyses in these compounds, the ESR linewidth above $T\rm{_V}$ is expected to be written as       
\begin{align}
\Delta H_{1/2} \propto \exp(\frac{E_{\rm{V}}}{k_{\rm{B}}T}),
\end{align}
where $E_{\rm{V}}$ is the activation energy of a free $Z_2$ vortex. A Monte Carlo simulation gives $T_{\rm{V}}=0.31JS^2$ and $E_{\rm{V}}=1.65JS^2$~\cite{Z2kawamura}, and thus the ratio $E_{\rm{V}}$/$T_{\rm{V}}$ becomes constant. We take $T_{\rm{V}}=8.5$ K from the anomaly of the magnetic susceptibility and the peak of the ESR linewidth and then obtain $E_{\rm{V}}=45.2$ K. Using this value, satisfactory agreement between experiment and calculation is obtained between 8.5 and 23 K as shown by the solid line in Fig. 4. In the neutron diffraction experiment, it was found that the correlation length in the triangular plane grows up to be about three lattice units at 23 K~\cite{neutron}. Our result indicates that a correlation length of about three lattice units is needed to observe the $Z_2$ vortex effect. The shift of the ESR resonance field below about 30 K is also thought to be caused by the development of the correlation length which is about two lattice units at 30 K~\cite{neutron}. 

For the case (d), it is expected that thermally excited free KT vortices fluctuate the spin system above a certain critical temperature $T_{\rm{KT}}$. Neglecting the temperature dependence of the vortex velocity, the temperature dependence of the ESR linewidth was reported to be proportional to a cube of the correlation length and given by~\cite{exp1990} 
\begin{align}
\Delta H_{1/2} \propto{\xi}^3=  \exp(\frac{3{\pi} }{2\sqrt{\frac{T}{T\rm{_{KT}}}-1}}).\label{KTdh}
\end{align}
The fit obtained with $T\rm{_{KT}}=8.5$ K cannot reproduce the gradual slope of the experimental result as shown by the dotted gray line in Fig. 4. The KT-type phase transition is accompanied by a divergence of the correlation length at $T_{\rm{KT}}$~\cite{KTmiyashita}. In NiGa$_2$S$_4$, no divergence was observed in such experiments. This is consistent with the fact that the obtained temperature dependence of the ESR linewidth deviates from Eq. ({\ref{KTdh}}).

In the low-temperature region (iii), the ESR linewidth shows a gradual decrease as the temperature is lowered and it remains a large value even at the lowest temperature. If there is a conventional phase transition to a LRO phase at about 8.5 K, the ESR linewidth is expected to decrease rapidly with decreasing temperature reflecting a development of uniform internal fields~\cite{VX2}. In the case of NiGa$_2$S$_4$, from the NQR measurements, Takeya $et$ $al.$ suggested that a gradual spin freezing occurs between 10 K and 2 K and that a inhomogeneous internal fields appear below 2 K~\cite{NMR}. A considerably large ESR linewidth at the lowest temperature must reflect a distribution of the inhomogeneous internal fields, which is consistent with the NQR result.

\subsection{ ESR resonance modes in the low-temperature region}
 Next, we discuss the frequency dependence of the ESR resonance fields of the strong signals at the lowest temperature. Figures 5(a) and 5(b) show the resonance fields at various frequencies for $H$ $\parallel$ $c$ and $H$ $\perp$ $c$, respectively. The broken lines represent paramagnetic-resonance lines with each $g$ factor, $g$$_{\parallel}$ = 2.08 and $g$$_{\perp }$ = 2.17. The ESR modes of the strong signal have a zero-field energy gap and approach the paramagnetic-resonance modes with increasing field for both field directions. These modes are reminiscent of the spiral spin resonance mode observed in a TLAFM with easy-plane anisotropy such as RbFe(MoO$_4$)$_2$~\cite{spiralmode}. In addition, it is known that some low-dimensional magnets in the temperature region where the SRO well develops exhibit resonance modes expected in the LRO phase~\cite{TMMC}. Therefore, we analyze the experimental results in terms of conventional spiral spin resonances. Considering strong third-neighbor and negligible second-neighbor exchange interactions, the calculation based on a mean-field approximation shows that the exchange constants are required to be $J$$_1$/$J$$_3$ = $-0.2$ and $J$$_2$ = 0 to realize the 57$^{\circ}$ spiral spin structure indicated by the neutron diffraction measurements. The $J$$_i$ ($i$ = 1, 2, 3) represent the first-, second-, and third-neighbor exchange interactions, respectively. Then, the spin Hamiltonian is written as,
\begin{align}
\mathcal {H} = J_1{\sum^{}_{<ij>}}\mib{S}_i{\cdot}\mib{S}_j+J_3{\sum^{}_{<lm>}}\mib{S}_l{\cdot}\mib{S}_m+D{\sum^{}_{i}}(\mbox{$S$}^{z}_{i})^2-g{\mu _B}{\sum^{}_{i}}\mib{S}_i{\cdot}\mib{H}\label{hamiltoni} 
\end{align}
where $D$ is the anisotropy of the easy-plane type ($D$ ${\textgreater}$ 0), ${\mu}$$_B$ the Bohr magneton, $\mib{H}$ the external magnetic field, the $z$-axis parallel to the $c$-axis, ${\langle}ij{\rangle}$ all the nearest neighbor pairs, and ${\langle}lm{\rangle}$ all the third-neighbor pairs. In our previous study, for simplicity, we assumed $J$$_1$=0, which results in a 120$^{\circ}$ spiral spin structure on the third nearest neighbor pairs in the $c$-plane~\cite{ESR}. In the present study, we take into account the finite $J$$_1$ to realize the 57$^{\circ}$ spiral spin structure, which is formed by 120 sublatices. Then, the free energy $F$ is expressed by the following form using the mean-field approximation,
\begin{align}
F = A{\sum^{}_{<ij>}}\mib{M}_i{\cdot}\mib{M}_j+B{\sum^{}_{<lm>}}\mib{M}_l{\cdot}\mib{M}_m+\frac{K}{2}{\sum^{120}_{i=1}}(\mbox{$M$}^{z}_{i})^2-{\sum^{120}_{i=1}}\mib{M}_i{\cdot}\mib{H}, 
\end{align}
where $A$, $B$, and $K$ is given by,
\begin{align}
A=\frac{120}{N}\frac{J_1}{(g{\mu _B})^2},\>\> B=\frac{120}{N}\frac{J_3}{(g{\mu _B})^2},\>\>K=\frac{120}{N}\frac{2D}{(g{\mu _B})^2},
\end{align}
and ${\mib{M}_i}$ is the $i$-th sublattice moment and given by  
\begin{align}
\mib{M}_i = \frac{N}{120}g{\mu _B}\mib{S}_i, 
\end{align}
where $N$ is the number of magnetic ions, and $\mib{S}_i$ is the spin on the $i$-th sublattice. We derive the resonance conditions by solving the equation of motion
\begin{align} {\partial}\mib{M}_i/{\partial}t=\gamma[\mib{M}_i\times\mib{H}_i],\label{toruku}
\end{align}
where $\gamma$ is the gyromagnetic ratio and $\mib{H}_i$ a mean-field applied on the $i$-th sublattice moment given by
\begin{align} 
\mib{H}_i=-{\partial}F/{\partial}\mib{M}_i.
\end{align}
To solve the equation of motion, we use a method applied for ABX$_3$ type antiferromagnets~\cite{tanaka}.
Assuming precession motions of the sublattice moments around those equilibrium directions, we utilize the following expressions, which represent the motion of the $i$-th sublattice moment,
\begin{align}
\mib{M}_i=({\Delta}M_{i\Acute{x}}\exp(i{\omega}t),{\Delta}M_{i\Acute{y}}\exp(i{\omega}t),|\mib{M}_i|).\label{henkanmae}
\end{align}
where ${\Delta}M_{i\Acute{x}},{\Delta}M_{i\Acute{y}}{\ll}|\mib{M}_i|$, and $\Acute{x}$, $\Acute{y}$ and $\Acute{z}$ are the principal axes of the coordinate system on each sublattice moment. The $\Acute{z}$-axis is defined to be parallel to the direction of the each sublattice moment, and the $\Acute{x}$ and $\Acute{y}$-axes are perpendicular to that. 

In the case of $H$ $\parallel$ $c$, all the sublattices are tilted from the easy plane with equivalent angle as shown by the schematic picture in Fig. 6(a). The angle ${\theta}$ between each individual sublattice moment and the external field is determined from the equilibrium condition ${\partial}F/{\partial}\theta=0$, and $\theta$ is given by
\begin{align}
\cos{\theta}=\frac{g_{\parallel}{\mu _B}H}{2S(D+3(J_1 +J_3)-2J_1\cos{\phi}-(J_1+2J_3)\cos{2\phi}-J_3\cos{4\phi})},
\end{align}
where $\phi=57$$^{\circ}$. Considering the change of ${\theta}$ under the external field, we obtain 120 resonance modes by solving Eq. ($\ref{toruku}$) numerically. If time dependent parts, namely transverse components, of the sublattice moments cancel each other and thus the total magnetization is not in motion, the resonance modes are, in principle, not observable. From the calculation of the total magnetization, we obtain two observable gapped modes, $h{\nu}_1$ and $h{\nu}_2$, and a gapless mode, $h{\nu}_3$ $=0$. The $h{\nu}_1$ calculated with the parameters $J$$_1$/$k\rm{_B}$ = $-4.56$ K, $J$$_3$/$k\rm{_B}$ = 22.8 K, $D$/$k\rm{_B}$ = 0.8 K exhibits good agreement between the experiment and the  calculation as shown in Fig. 5(a). The zero-field energy gap is evaluated to be about 13 K, which indicates a substantial anisotropy and decreases with increasing temperature to be about 10 K at $T_{\rm{v}}=8.5$ K. Since the $h{\nu}_2$ has a weak field dependence, it is expected to show a much broader resonance linewidth than $h{\nu}_1$, thus being hardly observed. In the case of $H$ $\perp$ $c$, we assume the distorted spin structure, in which spins lie in the easy plane as shown by the schematic view in Fig. 6(b). The distortion angles $\varphi\rm{_i}$ are determined so as to minimize the energy of the assumed structure by solving differential equations ${\partial}F/{\partial}\varphi_i=0$ ($i=1{\sim}120$), and we acquire them as numerical solutions. Then, through the same procedure as in the case of $H$ $\parallel$ $c$, we obtain two gapped modes, $h{\nu}_4$ and $h{\nu}_5$, and a gapless mode, $h{\nu}_6$ $=0$. The $h{\nu}_4$ mode shows good agreement with the experimental one using the same parameters for $H$ $\parallel$ $c$ as shown in Fig. 5(b). However, we did not observe the signals for $h{\nu}_5$. The reason is that the signal intensity of $h{\nu}_5$ mode is expected to be very weak due to the canceling of the transverse components of the sublattice moments, which is crucial for ESR transitions. This mode becomes soft around 45 T where one third of the saturation magnetization occurs.

Here, we briefly discuss the realization of a $Z_2$ vortex-induced topological transition on our system with a considerably large $D$ value. Quite recently, Kawamura derived the relation $D/k_{\rm{B}}{\le}T{\rm{_V}}(\xi/a)^{-2}$ which is a necessary condition for the $Z_2$ vortex-induced transition~\cite{kawamura_com}. In the case of a classical TLAFM model with only nearest-neighbor
interaction, the spin correlation length ${\xi}$ is estimated to be O(10$^3$)
at $T=T{\rm{_v}}$ so that even an extremely small $D$ would destroy the $Z_2$ vortex
transition. However, if the spin correlation length ${\xi}$ is supressed due to some
reason, the anisotropy is of the same order of $T_{\rm{{v}}}({\xi}/a)^{-2}$ or less,
and the $Z_2$ vortex-induced topological transition may still survive.

\subsection{Magnetization curves in the low-temperature region}

Using the parameters determined by the ESR analyses, we calculated the magnetization curves. For $H$ $\perp$ $c$, we take into account not only a spiral structure but also a so called umbrella structure with an out-of-plane spiral spin configuration. Then, from a mean-field theory, a first-order phase transition from the spiral structure to the umbrella structure is expected to occur at about 25 T. No such behavior, however, is observed and thus, spins must lie in the easy plane even above 25 T for $H$ $\perp$ $c$. The magnetizations are written as
\begin{align} 
M_{\parallel}=g_{\parallel}{\mu}_{B}S\cos{\theta}, \label{jikac}
\end{align}
\begin{align} 
M_{\perp} =g_{\perp}{\mu}_{B}S{\sum^{120}_{i=1}}\frac{\cos\varphi_i}{120}. \label{jikaab}
\end{align}
As shown in Fig. 6(a), the agreement between the experiment and the calculation is excellent in magnetic fields up to 63 T for $H$ $\parallel$ $c$. We reproduce the non-linear increase of magnetization for $H$ $\perp$ $c$ as shown in Fig. 6(b) by considering continuous changes of $\varphi_i$. The calculation suggests a first-order phase transition from a spiral spin structure to a fan structure at about 63 T for $H$ $\perp$ $c$, where the spins oscillate sinusoidally along the external field direction. However, the first order transition was not observed in our experiment. We cannot explain this disagreement within the framework of the mean-field theory.
In a Heisenberg TLAFM with the nearest-neighbor AF interactions, a phase transition from a spiral spin configuration to a collinear spin one, in which
the spins on the two sublattices are parallel and those on the other sublattice are antiparallel to the applied magnetic field in the
basal plane, is expected to occur at one third of the saturation magnetization due to quantum or thermal fluctuations~\cite{heisen_1/3,kawamura1/3}. A small anomaly of the magnetization observed at 44 T where the magnetization is close to one third of the saturation value, 0.72 ${\mu}$$_B$/ion, may be related to the 1/3 magnetization plateau discussed in the TLAFM. In our mean-field approximation corresponding to a classical calculation at $T=0$, the collinear configuration exists at a singular field~\cite{kawamura1/3}, and the calculated magnetization does not show any 1/3 magnetization plateau.

\subsection{Specific heat in the low-temperature region}
The above ESR resonance modes obtained by the mean-field theory correspond to the spin-wave excitations as a function of magnetic field at $q=0$, $\pm0.158$, where $q=ak/2\pi$ ($k$ is a wavevector and $a$ is the lattice constant of the triangular plane). This suggests that spin-wave-like excitations develop in the basal plane at low temperatures despite the absence of the LRO in NiGa$_2$S$_4$. In Fig. 7, we show spin-wave dispersion energies, $h{\nu}_{\rm{_k}}$, at $H$=0 T and $H$=7 T for $H$ $\parallel$ $c$ calculated by means of a conventional spin-wave theory with parameters obtained by our ESR analyses. We find a $k$-linear dispersion around $k\sim 0$, which remains gapless even in a magnetic field up to 7 T. Furthermore, the spin-wave velocity $v\rm{_s}$, which corresponds to the slope of the $k$-linear dispersion, is almost constant in magnetic fields up to 7 T. Actually, we obtain $v\rm{_s}=1045$ m/s at 7 T and $v\rm{_s}=1049$ m/s at 0 T, which are almost equivalent to each other and are nearly the same value as that evaluated from a neutron scattering experiment.~\cite{neutron} This almost field-independent dispersion must cause $T^2$-dependent specific heat below about 10 K. Therefore, we evaluate the specific heat in the low-temperature region. The specific heat is derived from the derivative of the average thermal energy written as
\begin{align} 
C={\partial}E/{\partial}T.\label{C}
\end{align}
From the classical spin-wave theory, the $E$ is given by  
\begin{align} 
E={\int}^{\infty}_0\frac{h{\nu}_{\rm{_k}}}{\exp(\frac{h{\nu}_{\rm{_k}}}{k{\rm{_B}}T})-1}D(k)dk,\label{E}
\end{align}
where $D(k)$ is a density of states. In the present triangular lattice model, it is given by, 
\begin{align} 
D(k)=\frac{a^{2}N{\rm{_A}}}{2{\pi}}k.\label{D}
\end{align}
where $N\rm{_A}$ is the Avogadro number. Below about 10 K, we only consider the linear dispersion around $k\sim 0$ because the energy gap was evaluated to be about 13 K, and the spin-wave dispersion obeys the relation $h{\nu}_{\rm{_k}}{\approx}v{\rm{_s}}k$. Accordingly, the specific heat obtained from the above relations is calculated as
\begin{align} 
C=\frac{3{\zeta}(3)}{{\pi}}N{{\rm{_A}}}k_{\rm{B}}^{3}(\frac{aT}{hv{\rm{_s}} })^2,\label{calC}
\end{align}
where ${\zeta}(3)=1.202$, $h$ Planck's constant. Although we confirm that the obtained $v\rm{_s}$ is close to the value evaluated from the recent neutron diffraction measurement~\cite{neutron}, the calculated $C$ value is nearly 36 times smaller than the experimental one. For quantitative agreement with the experimental specific heat, the value of $v\rm{_s}$ is required to be nearly six times~\cite{teisei} smaller than that obtained from our ESR analyses. We consider that the quantitative disagreement in the present spin-wave model may come from some kind of effect, for example, anomalous structure factor of the excitation mode or quadrupolar correlations stabilized by a biquadratic interaction~\cite{AFQ,newquadra}. For the latter, Kawamura and Yamamoto predicted by means of their Monte Carlo simulations that any type of biquadratic interaction could couse a $Z_2$ vortex-induced topological transition~\cite{kawamuraNiGaS}.

\section{Conclusions} 
We performed high field ESR and magnetization measurements on single crystals of the triangular lattice antiferromagnet NiGa$_2$S$_4$. We find three distinct temperature regions: (i) above 23 K, (ii) between 23 K and 8.5 K, and (iii) below 8.5 K. In the high-temperature region (i), the temperature dependence of the ESR linewidth indicates a development of the SRO of the 2D Heisenberg AFM. In the intermediate temperature region (ii), the development of the SRO is suppressed by thermally excited $Z_2$ vortices and the temperature dependence of the ESR linewidth is proportional to the exponential function with the activation energy of the free $Z_2$ vortex toward 8.5 K. In the low-temperature region (iii), the frequency dependence of the ESR resonance fields and the magnetization processes are explained to a large extent by the conventional mean-field theory. Furthermore, we derive a gapless linear dispersion relation for $H$ $\parallel$ $c$ from the conventional spin-wave theory and explain qualitatively the field-independent specific heat with $T^2$ dependence. The origin of the large quantitative disagreement between experiment and calculation based on classical spin models suggest significant contribution from quantum effects, possibly, due to quadrupolar correlations. Although there are some distinctions from the conventional spin-wave model, the observed properties, explained by the spin-wave excitations in spite of the absence of LRO, are consistent with those expected in the $Z_2$ vortex-induced phase~\cite{kawamuraNiGaS}. Consequently, unconventional properties of NiGa$_2$S$_4$ in the low-temperature region strongly suggest that a vortex-induced topological transition takes place at 8.5 K.

\section*{Acknowledgements} 
We thank H. Kawamura and T. Okubo for valuable discussion and critical reading of the manuscript and S. Todo for advice about numerical calculations. This work was supported by Grants-in-Aid (No.17072005, No.19052003, and No.
20340089, No. 21684019), and by the 21st Century COE (Project No. G17) and Global COE (Project No. G10) Programs from the
Ministry of Education, Science, Sports, Culture and Technology in
Japan.

\begin{figure}[t]
\begin{center}
\includegraphics[width=10cm]{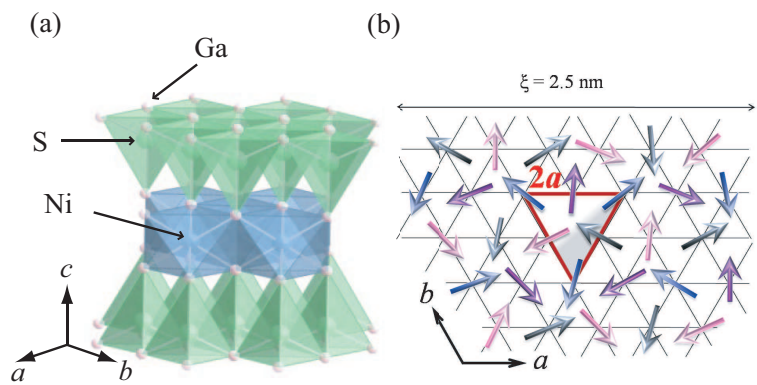}
\end{center}
\caption{(Color online) (a)Structure of the unit slab of NiGa$_2$S$_4$ consisting of two GaS layers and one NiS$_2$ layer stacking along the $c$-axis. (b)Short range correlated spiral spin structure of NiGa$_2$S$_4$ at low temperature region. The $ab$-plane components of spins rotate with an angle about 57$^{\circ}$.}
\label{f1}
\end{figure}

\begin{figure}[t]
\begin{center}
\includegraphics[width=13cm]{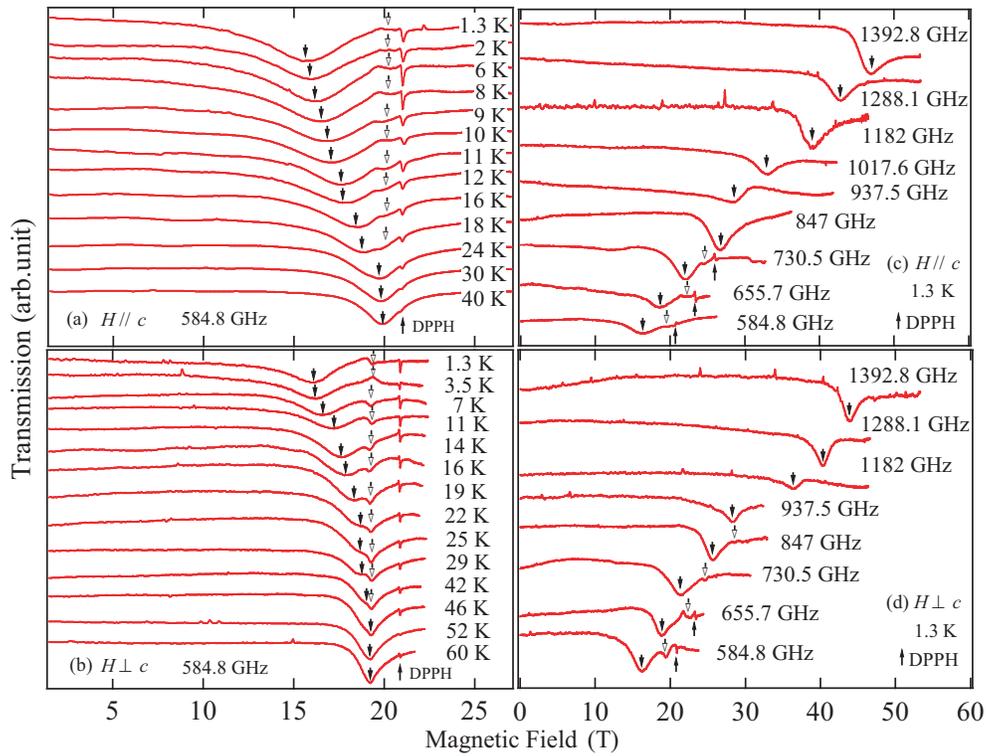}
\end{center}
\caption{(Color online) (a), (b) Temperature dependence and (c), (d) frequency dependence of ESR absorption spectra of NiGa$_2$S$_4$ at 584.8 GHz and at 1.3 K for $H$ $\parallel$ $c$ ((a),(c)) and $H$ $\perp$ $c$ ((b),(d)). The closed and open arrows indicate the broad large signal and the weak one, respectively. The sharp signals at higher field come from an ESR standard sample of DPPH for correction of the magnetic field.}
\label{f2}
\end{figure}

\begin{figure}[t]
\begin{center}
\includegraphics[width=7cm]{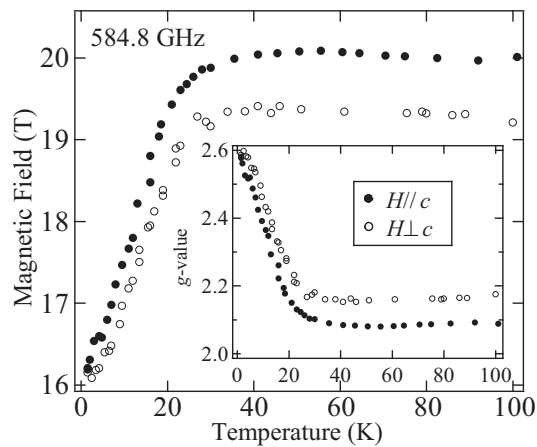}
\end{center}
\caption{(Color online) Temperature dependence of the resonance field of the strong ESR signals in NiGa$_2$S$_4$ at 584.8 GHz. Open and closed circles represent the resonance fields for $H$ $\parallel$ $c$ and $H$ $\perp$ $c$, respectively. The inset shows $g$-values converted from the resonance fields.}
\label{f3}
\end{figure}

\begin{figure}[t]
\begin{center}
\includegraphics[width=7cm]{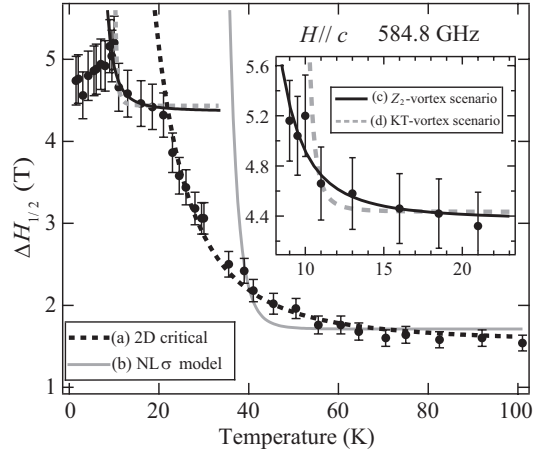}
\end{center}
\caption{(Color online) Temperature dependence of the full width at half maximum of the broad ESR spectra in NiGa$_2$S$_4$ for $H$ $\parallel$ $c$ at 584.8 GHz. Black broken, gray solid, black solid, and gray broken lines represent the temperature evolution of the linewidth calculated for (a) 2D critical behavior, (b) non-linear sigma (NL$\sigma$) model, (c) $Z_2$-vortex scenario, and (d) Kosterlitz-Thouless (KT) vortex scenario, respectively. The inset shows the expansion of the temperature region between 8 K and 23 K.}
\label{f4}
\end{figure}

\begin{figure}[t]
\begin{center}
\includegraphics[width=8cm]{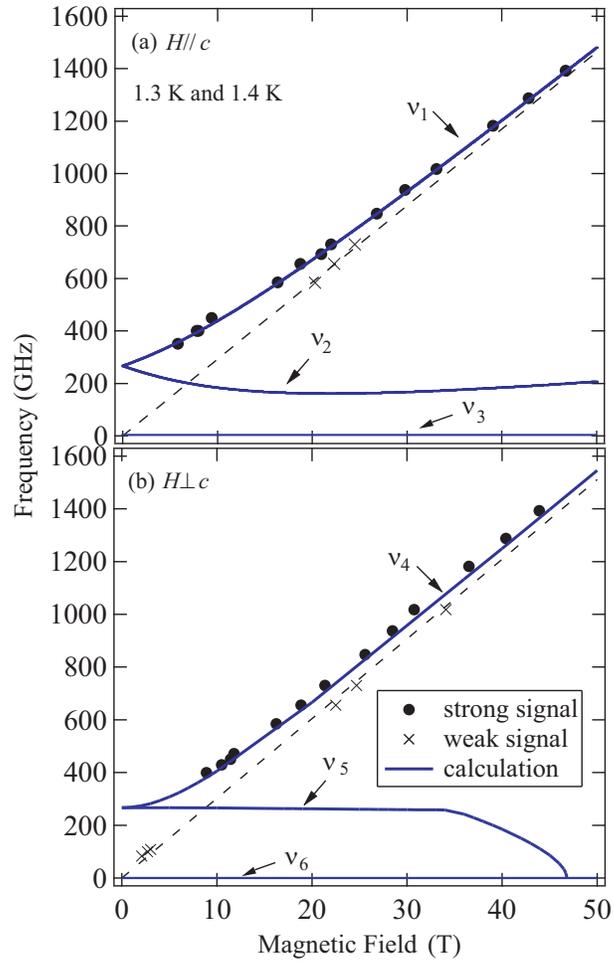}
\end{center}
\caption{(Color online) Frequency-field plot of the resonance fields of NiGa$_2$S$_4$ measured at 1.3 K in pulsed magnetic fields and at 1.4 K in static magnetic fields for (a) $H$ $\parallel$ $c$ and (b) $H$ $\perp$ $c$. Closed circles and crosses denote strong and weak signals, respectively. Solid and broken lines represent calculated resonance modes and the paramagnetic-resonance lines, respectively.}\label{f4}
\end {figure}

\begin{figure}[t]
\begin{center}
\includegraphics[width=8cm]{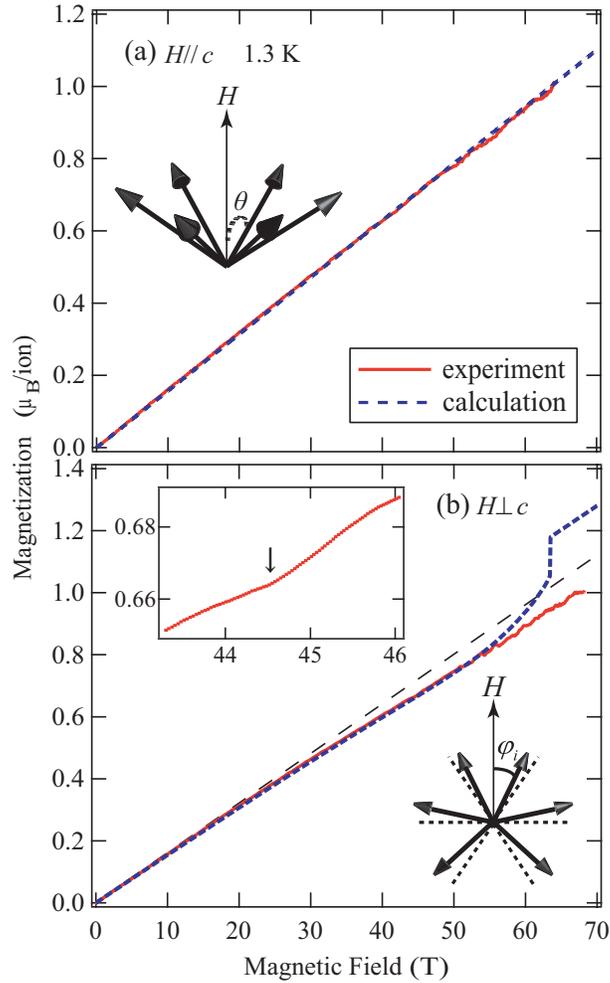}
\end{center}
\caption{(Color online) Magnetization curves of NiGa$_2$S$_4$ at 1.3 K for (a) $H$ $\parallel$ $c$ and (b) $H$ $\perp$ $c$. Solid and blue broken lines are experimental and calculated magnetization curves, respectively. The black thin broken line is an extension of the liner increase at low-field region. The inset of (b) shows the vicinity of the anomaly for $H$ $\perp$ $c$. The illustration in each figure shows a schematic view of spin configuration under an external magnetic field for each direction. }\label{f5}
\end {figure}

\begin{figure}[t]
\begin{center}
\includegraphics[width=7cm]{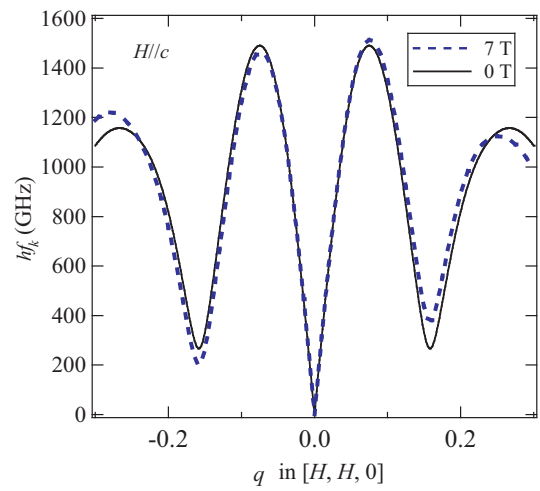}
\end{center}
\caption{(Color online) Calculated spin-wave dispersion relation of NiGa$_2$S$_4$. Solid and broken lines represent the relations at $H=0$ T and $H=7$ T for $H$ $\parallel$ $c$, respectively.}\label{f7}
\end {figure}


\begin{thebibliography}{99}
\bibitem{teisei} In Ref. \ref{ESR}, we reported "three times" instead of "six times" in the present paper 
by mistake.\label{teisei}
\end{thebibliography}
\end{document}